\documentstyle[11pt]{article}
\newcommand{\im}{{\rm Im}~}
\newcommand{\re}{{\rm Re}~}
\newcommand{\wsq}{{|W_+|}^2}
\newcommand{\ws}{{|{\cal W}|}^2}
\newcommand{\ww}{{\cal W}}
\newcommand{\bb}{{\cal B}}
\newcommand{\bbar}{\overline{{\cal B}}}
\newcommand{\ebar}{\overline{\varepsilon}}
\newcommand{\th}{\mbox{$\vartheta_1$}}
\newcommand{\be}{\begin{equation}}
\newcommand{\ee}{\end{equation}}
\newcommand{\tr}{\tau_{\mbox{${\scriptscriptstyle R}$}}}
\newcommand{\ti}{\tau_{\mbox{${\scriptscriptstyle I}$}}}
\newcommand{\bel}[1]
{\begin{equation} \mbox{$\label{eq:#1}$}}
\newcommand{\beal}[1]
{\begin{eqnarray} \mbox{$\label{eq:#1}$}}
\newcommand{\half}{{{1}\over{2}}}

\newcommand{\okl}[1]{\omega_{kl}^{#1}}
\newcommand{\osq}{{|{\omega_{kl}}|}^2}
\newcommand{\dd}[2]{{{\partial #1}\over{\partial #2}}}
\newcommand{\ddd}[2]{{{{\partial}^2 #1}\over{\partial {#2}^2}}}
\newcommand{\dddd}[3]{{{{\partial}^2 #1}\over
{\partial #2 \partial #3}}}
\newcommand{\klsum}{\sum_{k,\, l=-\infty}^\infty~}
\newtheorem{theorem}{Theorem}
\newcommand{\qed}{\rule{1.1ex}{0.7em}}
\newcommand{\comma}{\hspace{.3cm} {\rm ,}}
\newcommand{\period}{\hspace{.3cm} {\rm .}}
\newcommand{\ie}{i.\frenchspacing e.\nonfrenchspacing \ }
\newcommand{\eref}[1]{Eq.\,(\ref{eq:#1})}
\newcommand{\fotnot}[1]{\frenchspacing \footnote{#1} \nonfrenchspacing}

\def\themonth{\ifcase\month\or
January\or February\or March\or April\or May\or June\or
July\or August\or September\or October\or November\or December\fi}
\newcommand{\YCTP}[2] {\mbox{} \\[.15cm]
\noindent{\large \themonth\ 1992\hfill YCTP\,--\,P#1\,--\,#2} \vspace{.5cm}}
\newcommand{\mytitle}[1]
{\begin{center} {\LARGE #1} \end{center}}
\newcommand{\myauthor}[1]
{\renewcommand{\thefootnote}{\fnsymbol{footnote}}
\begin{center} {\large #1} \end{center}
\renewcommand{\thefootnote}{\arabic{footnote}}
\setcounter{footnote}{0} \vspace{1cm}}
\newcommand{\email}[1]{\footnote{#1}}

\begin{document}
\setlength{\baselineskip}{24pt}
\YCTP{11}{92}
\mytitle{GINZBURG-LANDAU THEORY OF THE ELECTROWEAK PHASE
TRANSITION AND ANALYTICAL RESULTS}
\myauthor{Ola T\"{o}rnkvist\email{Electronic address TORNKVIST@YALEHEP}
\\[-.4cm]
Department of Physics, Yale University, New Haven, CT 06511, U.S.A.}

\noindent
{\bf Abstract.} The phase transition of the electroweak vacuum induced
by a strong magnetic field is examined, and a
connection is made with the Ginzburg-Landau theory of
type-II superconductivity.
For solutions of the exact
nonlinear field equations of the electroweak theory
with lattice periodicity in
directions perpendicular to the magnetic field, it is proven
that, likewise, each
lattice cell must enclose an integer number of quanta of magnetic flux.
Close to the lower critical magnetic field, a
perturbative method developed by MacDowell and the author is used to
study properties of the lattice solutions.
Analytical expressions for observables are obtained in
terms of a complex parameter $\tau$ specifying the lattice and
it is shown that the triangular
Abrikosov solution constitutes a local minimum of the energy
provided $M_H > M_Z$.

\newpage
\section{Introduction}
\label{sec-intro}

In the cosmological scenario, the electroweak transition between the
symmetric
{\nolinebreak $SU(2) \otimes U(1)_Y$}
 and the broken ${U(1)}_{EM}$ phases
results from a temperature dependence in the coefficients of the
Higgs potential. The transition occurs at a critical temperature
$T_c$ and may involve the coexistence of phases, depending on the order of
the transition.

It has been shown by Ambj{\o}rn and Olesen \cite{AO3,AO4-AO5} that the phase
transition can be induced at {\em zero temperature} by a large
magnetic field. Their analysis was done with particular values of the
coupling constants, for which the nonlinear field equations simplify,
corresponding to $M_H = M_Z$ where $M_H$ is the Higgs-boson mass. The
transition from the broken phase with Higgs field
$\phi \equiv {\phi}_0$
to the symmetric phase with $\phi \equiv 0$ was found to take place
gradually as the magnetic field increases from $B_{c2} = M_W^2 / e$
to $B_{c1} = M_Z^2 / e$.\fotnot
{With this choice of labels for the critical magnetic fields,
the phase transition at $B_{c2}$ is qualitatively similar to the
well studied transition at $B_{c2}$ in a type-II superconductor.
In Refs.\,\cite{AO3,AO4-AO5} the opposite labels were used.}
Between these values the field equations for
 $A_{\mu}$,
$Z_{\mu}$, $W_{\mu}$, and $\phi$ admit solutions with
lattice periodicity in directions perpendicular to the magnetic
field \cite{Yang}.
The solutions have been obtained numerically and studied
in the range
$B_{c2} < B < B_{c1}$ \cite{AO3,Varsted}.
The emerging vacuum structure resembles that of the mixed state
in a type-II
superconductor \cite{StJames} with the
order parameter given
by a density
of W-boson pairs forming a zero-charge condensate.
The distinctive feature is that the
electroweak vacuum is paramagnetic, \ie the magnetic field is
enhanced by the W condensate, while a superconductor is diamagnetic.

The phase transition resolves an old problem of the vacuum instability
\cite {AO0,Ska78} of
the Weinberg-Salam model for magnetic fields of
magnitude $B_{c2} =
M_W^2 / e \approx 10^{20}$ Tesla and beyond.
The instability is a consequence of the large magnetic moment of the W
bosons, which, when aligned with the magnetic field, can lower the energy
sufficiently to make the W condensate energetically more favorable
than the trivial vacuum. The W pairs also couple to and
act as sources for
nontrivial Z and Higgs fields.

Explicit analytical solutions to the full, nonlinear problem are unknown,
even for the case $M_H = M_Z$ considered in Refs.\,\cite{AO3,AO4-AO5}.
Approximate solutions can be obtained, however, in certain limits.
For $B$ near the upper critical field $B_{c1}$ $(\phi \approx 0)$ a
perturbative solution was recently given by Olesen \cite{O7}.
In the vicinity of the lower critical field $B_{c2}$ $(\phi \approx
{\phi}_0)$ a perturbative method was first suggested by Skalozub
\cite{Ska87}, who found that, in this regime, the solutions with
lattice symmetry coincide with the Abrikosov solutions \cite{Abrikosov}
for a type-II superconductor near $B_{c2}$. In Skalozub's derivation,
the interactions involving the $Z$ and $\phi$ fields were approximated
with a local quartic interaction.

In a recent paper \cite{MT1992a}, MacDowell and T\"{o}rnkvist
applied a perturbative method to the full Weinberg--Salam model
for general values of the
coupling constants and solved the field equations exactly to lowest
order in $(B - B_{c2})$. In this approach, the interactions mediated by $Z$
and $\phi$ were accounted for in an effective, {\em nonlocal\/} quartic
interaction. An investigation showed that, for $M_H > M_Z$, the
triangular\fotnot{In Ref.\,\cite{MT1992a} it was denoted ``hexagonal''.}
Abrikosov lattice solution
represents the energetically most favorable configuration.

In this paper we shall derive analytical expressions for physical
observables, such as the energy density,
as a function of the geometry of the lattice solutions and of $(B-B_{c2})$.
The geometry is specified by a complex parameter $\tau$.
In the complex plane, $\tau$ and the real number $1$ span
a fundamental parallelogram for an
arbitrary two-dimensional simple lattice.

\newpage
The paper is organized as follows.
In section \ref{sec-EWGL} the perturbative method developed in
Ref.\,\cite{MT1992a} is reviewed. Physical
quantities are shown to depend on the function $\ws$ where $\ww$ is
the perturbative solution for the W field.
A general theorem of magnetic flux quantization is presented.
For any solution of the nonlinear electroweak field
equations where $U(1)_{EM}$ gauge-invariant quantities
possess a lattice symmetry in directions perpendicular to the
magnetic field, it is shown that each lattice cell must enclose an
integer number of quanta of abelian magnetic flux.

Section \ref{sec-ExpAS} is devoted to mathematical properties
of the Abrikosov solutions. A Fourier expansion of $\ws$
is derived and provides a fast-convergent representation with a
simple dependence on the lattice parameter $\tau$. This representation
is used in section \ref{sec-AnalExp} to derive explicit expressions for
physical observables in terms of $\tau$. It is shown analytically
that the triangular lattice solution constitutes a local
minimum of the energy, provided $M_H > M_Z$.

\section{Electroweak Ginzburg-Landau Theory}
\label{sec-EWGL}

In the unitary gauge, the Weinberg-Salam lagrangean density
\cite{Svartholm} leads to coupled equations for $A_{\mu}$,
$Z_{\mu}$, $W_{\mu}$ and the real Higgs field $\phi$ (see
Ref.\,\cite{MT1992a}~for details). The equations for $W_{\mu}$
and $A_{\mu}$ are
\bel{W}
D^{\mu} F_{\mu \nu}= i g {\left[ \cos \theta Z_{\mu \nu} +
\sin \theta f_{\mu \nu} - i g (W_{\mu}^{\dag} W_{\nu} -
W_{\nu}^{\dag} W_{\mu}) \right]} W^{\mu} - {{1}\over{2}}
g^2 {\phi}^2 W_{\nu} \comma
\ee
\bel{A}
{\partial}^{\mu} f_{\mu \nu}= i e {\left[
{\partial}^{\mu}(W_{\mu}^{\dag} W_{\nu} -
W_{\nu}^{\dag} W_{\mu}) + W^{\mu \dag} F_{\mu \nu} -
W^{\mu} F_{\mu \nu}^{\dag}
 \right]} \comma
\ee

\noindent
where
$D^{\mu} = {\partial}^{\mu} + i g (A^{\mu} \sin \theta +
Z^{\mu} \cos \theta )
$,~$F_{\mu \nu} = D_{\mu} W_{\nu} - D_{\nu} W_{\mu}$,
{}~$f_{\mu \nu} = {\partial}_{\mu} A_{\nu} - {\partial}_{\nu} A_{\mu}$,
{}~$Z_{\mu \nu} = {\partial}_{\mu} Z_{\nu} - {\partial}_{\nu} Z_{\mu}$
and $e = g \sin \theta$.
The field $W^{\mu}$ is subject to the constraint \cite{MT1992a}
\bel{Wconstraint}
D^{\nu} W_{\nu} = W_{\nu} {\left( {{i g}\over {\cos \theta}} Z^{\nu}
- {\partial}^{\nu} \ln {\phi}^2 \right)} \period
\ee

\newpage
We shall investigate static configurations of fields
which include a magnetic field $\vec{B}$ in the $\hat{z}$ direction
and where all fields are independent of the $z$ coordinate. One can
show that it is sufficient to consider components of the vector
potentials in the $\hat{x}$ and $\hat{y}$ directions and that no
inconsistent $\hat{z}$ or time components develop dynamically.
The resulting problem is then two-dimensional.

Define the spin projection states $W_{\pm}=(W_1 \mp i W_2)/{\sqrt{2}}$.
For $B \leq B_{c2} = M_W^2 / e$, Eqs.\ (\ref{eq:W})
and (\ref{eq:A}) have the trivial solution $W_{\mu}=0$, $B = f^1{}_2 \equiv$
const., $\vec{A}= (\vec{B} \times \vec{r}~) / 2$,
$\vec{Z}=0$ and $\phi \equiv {\phi}_0$.   When the magnetic field $B$
exceeds $B_{c2}$, perturbations about zero in
the linear combination $W_+$ appear to become tachyonic
in the field equation, \eref{W}, because of the spin interaction term
$i g \sin \theta f_{\mu \nu} W^{\mu}$. The trivial vacuum thus becomes
unstable with respect to
the production of pairs of W bosons with magnetic moments
oriented along the magnetic field.
Stability is restored by the cubic term in \eref{W}
and by the back reactions
$\vec{Z}$, $\phi - {\phi}_0$, $B - B_{c2}$, and
$\vec{A} - {\vec{A}}_{c2}$ for which $W_+$ acts as a source.
In Ref.\,\cite{MT1992a} it was shown that these back
reactions are all of order $\wsq$.
In particular, $B$ in this order obeys the linear relation
\bel{Blin}
B - e \wsq \equiv H \comma
\ee
\noindent
where $H$ is a spatially uniform field.

Using \eref{Wconstraint}, one can also
show \cite{MTGauss}  that the
suppressed component $W_-$ is of order ${|W_+|}^3$.
Therefore, for an  average magnetic field
$\bar{B}$ above and near $B_{c2}$,
we can write the equation for zero-energy eigenstates $W_+$ as
follows:
\bel{Wlin}
{\left[ - {(\nabla - i e \vec{\bar{A}})}^2 + M_W^2 - 2 e \bar{B}
 + {\cal O} (\wsq) \right]} W_+ = E^2 W_+ = 0 \comma
\ee
\noindent
where $\vec{\bar{A}}$ is merely a  notation for a vector potential
such that
$\nabla \times \vec{\bar{A}} = \bar{B} \hat{z}$.

\newpage
We recognize \eref{Wlin} as a generalization of the
first Ginzburg-Landau equation of type-II superconductivity.
That equation is the special case where ${\cal O} (\wsq)$ is
a positive constant times $\wsq$ and corresponds to a hamiltonian
density with an effective local quartic interaction. In
contrast, the effective hamiltonian of
the electroweak problem is nonlocal.
It was derived in Ref.\,\cite{MT1992a}
by expanding the exact hamiltonian
up to second order in $\wsq$ including
back reactions on the fields.
The resulting
effective hamiltonian is
\bel{Hamilton}
{\cal E}(\vec{r}~) = {{1}\over{2}} B^2 - (e B - M_W^2) \wsq +
 {{1}\over{2}} g^2
\left[~{\sin}^2 \theta~{|W_+|}^4 + M_W^2 U(\vec{r}~) \wsq \right] \comma
\ee
\noindent
where
\bel{U}
U(\vec{r}~)= {{1}\over{2 \pi}}
\int d^2 r' \left[ K_0(M_Z|\vec{r} - {\vec{r}} \, '|) -
K_0(M_H|\vec{r} - {\vec{r}} \, '|) \right] {|W_+({\vec{r}} \, ')|}^2
\comma
\ee
\noindent
$K_0$ is a Bessel function and $M_Z$,
$M_H$ are the masses of the Z and Higgs bosons.

The interaction involving $U(\vec{r}~)$
can be interpreted as an effective long-range interaction  between W pairs,
mediated by Z and Higgs bosons.
For $B > B_{c2}$, \eref{Hamilton} shows that the minimal energy
 occurs for a non-zero $W_+$ field.
 In the case
$M_H < M_Z$, one can find configurations which yield a negative
quartic term. It is then necessary to go to higher orders of perturbation
theory, which is beyond the scope of this paper.
There are indications that an electroweak generalization of the Nielsen-Olesen
vortex may be the stable solution in this mass regime \cite{Vachaspati}.
This cannot be verified with a perturbative approach.

If $M_H \geq M_Z$,
stability is ensured
by the quartic interaction.
Then there exist perturbative solutions of \eref{Wlin}
for which $\wsq$ is periodic on a
fine-grain lattice of parallelograms and uniform on a macroscopic
scale. They are known, from previous work in superconductivity,
as Abrikosov flux lattice solutions \cite{Abrikosov}.
\newpage \noindent
 In this paper,
the solutions will be reconsidered in the context of the
electroweak theory and physical quantities will be expressed as
functions of the geometry of the lattice.
This geometry is specified by the two lattice vectors
$a \hat{x}$ and $a \vec{\tau}$ or, equivalently, by the complex numbers
$a$ and $a \tau$,~$a \in \Re$, with the correspondence
 $\vec{\tau} = \re \tau~ \hat{x} +
\im \tau~ \hat{y}$. In the complex picture,
$\wsq$ is referred to as {\em doubly periodic} with
periods $a$ and $a \tau$.

Introduce the notation $\tr = \re \tau$ and $\ti = \im \tau$.
 For a given lattice, the cell side $a$ and area
$A= a^2 \ti$ are dynamically determined by the value of the
average magnetic field $\bar{B}$ through the following
theorem, which is valid also for nonperturbative solutions.
\begin{theorem}[Flux Quantization Condition]\fotnot
{The theorem has previously been
shown to hold for perturbative solutions
\cite{MT1992a} and, nonperturbatively, for the special case of
the electroweak theory
where $M_Z = M_H$ \cite{AO2}. This proof was constructed in
collaboration with MacDowell.}
For field configurations where ${U(1)}_{EM}$ \linebreak
 gauge-invariant
quantities, such as $f_{12}$,
$Z_{\nu}$, $\phi$, $|W_+|$, $|W_-|$ and
$W_- / W_+$, are doubly periodic,
the abelian magnetic flux that penetrates each
lattice cell of area $A$ is quantized and restricted to the values
$\bar{B} A = 2 \pi k / e$ , where the integer $k$ is the
common winding number of the phases of $W_+$ and $W_-$.
\end{theorem}
\noindent
Proof:~~~Through integration by parts, \eref{A} can be
written
\beal{proof}
{\partial}^{\mu} f_{\mu \nu} & = & 2 i e
{\partial}^{\mu}(W_{\mu}^{\dag} W_{\nu} -
W_{\nu}^{\dag} W_{\mu}) \nonumber \\*
& - & i e \left[
{(D^{\mu} W_{\mu})}^{\dag} W_{\nu} - W_{\nu}^{\dag} (D^{\mu} W_{\mu})
\right] \nonumber \\*
& + & \underbrace{i e \left[
{(D_{\nu} W^{\mu} {)}^{\dag} W_{\mu} -
 W^{\mu \dag} (D_{\nu} W_{\mu})}
\right]}_{j_{\nu}^{{\rm ~top}}} \period
\end{eqnarray}

\noindent
Let \hfill
$W_+ = |W_+| e^{i {\chi}_+}$ \hfill and \hfill
$W_- = |W_-| e^{i {\chi}_-}$.\hfill
The last term on the right hand side then
\newpage
\noindent
becomes
\beal{phase}
\nopagebreak
{j_{\nu}^{{\rm top}}} & = & \mbox{} - 2 e^2 (\wsq + {|W_-|}^2 )
{}~{\left[ A_{\nu}
 + {{1}\over{2 e}} {\partial}_{\nu} ({\chi}_+ + {\chi}_-)
\right]} \nonumber  \\*
& & \mbox{} - 2 e g \cos \theta (\wsq + {|W_-|}^2 ) Z_{\nu}
 - e (\wsq - {|W_-|}^2 ) {\partial}_{\nu} ({\chi}_+ - {\chi}_-) \period
\end{eqnarray}
\noindent
The vector ${\partial}_{\nu} ({\chi}_+ - {\chi}_-)$ is invariant under
lattice translation by virtue of the periodicity of $W_- / W_+$.
Consequently, its line integral around the
boundary of a parallelogram vanishes and, with the requirement that fields
be single valued,
\bel{nequal}
\oint {\partial}_{\nu} {\chi}_+ dx^{\nu} =
\oint {\partial}_{\nu} {\chi}_- dx^{\nu} = 2 \pi k,~~
k~{\rm integer.}
\end{equation}

\noindent
Thus the phases of $W_+$ and $W_-$ have the same winding number $k$.
Using \eref{Wconstraint}
and, again, the periodicity of $W_- / W_+$, one can show that
the middle term of \eref{proof} is
an invariant vector under translation.
When the expression
for ${j_{\nu}^{{\rm top}}}$, \eref{phase}, is substituted into
\eref{proof}, the latter
divided by $2 e^2 (\wsq + {|W_-|}^2 )$
can be rearranged in the form
\bel{Achi}
\mbox{} - A_{\nu} =
 {{1}\over{2 e}} {\partial}_{\nu} ({\chi}_+ + {\chi}_-) + j_{\nu}^
{{\rm inv}} \comma
\ee
\noindent
where $j_{\nu}^{{\rm inv}}$ is an invariant vector under lattice translation.
Therefore the integral $\oint j_{\nu}^{{\rm inv}} dx^{\nu}$ around
the boundary of a parallelogram vanishes, and we get
\bel{winding}
\hspace{2.5cm} {\rm flux} = \mbox{} - \oint A_{\nu} dx^{\nu} =
 {{1}\over{2 e}} \oint {\partial}_{\nu} ({\chi}_+ + {\chi}_-) dx^{\nu}
= {{2 \pi k}\over{e}} \hspace{2.5cm} \qed
\ee
\\
\noindent
The periodicity condition on $W_- / W_+$ is required
for the theorem to hold. It emerges naturally,
if one assumes that the
phases acquired under a lattice translation correspond
to a gauge transformation,
where the vector potential is form-invariant but expressed about the
translated origin  \cite{MTGauss}.
We remind ourselves that $W_-$ and $W_+$ have the same ${U(1)}_{EM}$ charge
and that $W_- / W_+$ therefore is a gauge invariant quantity.

\newpage
{}From the flux quantization condition we find that
the side $a$ is determined by the relation
\bel{adef}
a^2 = {{2 \pi k}\over{e \bar{B} \ti}} \period
\ee
It is convenient to redefine
the problem in terms of coordinates, where
the sides of a lattice cell
have lengths $1$ and $|\tau|$. The lattice is then specified by
the sole parameter $\tau$, and we can impose $\ti > 0$ with no
lack of generality.

Define the dimensionless quantities $\cal B$, $\vec{\rho}$,
$V({\vec{\rho}})$, ${\cal W}$, $\varepsilon$, and $\kappa$ by
\bel{scale}
{\cal B} = {{e B}\over{ M_W^2}}, \hspace{.3cm} \vec{r}= a \vec{\rho}
,\hspace{.3cm} V(\vec{\rho}~) = g^2 U(\vec{r}~)
,\hspace{.3cm} W_+ = {{M_W}\over{e}} {\cal W}, \hspace{.3cm}
{\cal E} ={{M_W^4}\over{e^2}} \varepsilon,
\hspace {.4cm} \kappa = {{k \pi}\over{\ti }} \period
\ee
\noindent
With these units, the effective hamiltonian, \eref{Hamilton},
becomes
\bel{Enew}
\varepsilon (\vec{\rho}~) = {{1}\over{2}} {\cal B}^2 - ({\cal B} - 1)
\ws + {{1}\over{2}}
\left[~{|{\cal W}|}^4 + V(\vec{\rho}~) \ws \right]
\ee
\noindent
and the critical magnetic field is ${\cal B}_{c2} = 1$.
{}From \eref{Blin} it follows that, in the new units,
${\cal B} - \ws \equiv \bbar - \overline{\ws} \equiv h$ is a uniform
field. Substituting this relation
into the effective hamiltonian we obtain the
space-averaged energy density in terms of the average magnetic field
$\bbar$.
\bel{Eave}
\ebar = {{1}\over{2}} {\bbar}^2 - (\bbar - 1)
\overline{\ws} + {{1}\over{2}}
\left[~{( \overline{\ws} )}^2 + \overline{V(\vec{\rho}~) \ws} \right]
\period
\ee
\noindent
The lattice geometry does not fix the overall normalization of the solution
$\ww$. For a given geometry, the normalization can be determined by
 minimizing the energy with
a fixed average magnetic field $\bbar$. The resulting condition is
\bel{Wnorm}
- (\bbar - 1) \overline{\ws} +
\left[~{( \overline{\ws} )}^2 + \overline{V(\vec{\rho}~) \ws} \right]
= 0 \period
\ee
\newpage
\noindent
Physical quantities can then be expressed in terms of $\bbar$ and the
quantity
\bel{beta}
\beta = {\textstyle {\overline{V(\vec{\rho}~) \ws}}\over
{\textstyle {( \overline{\ws} )}^2}} \comma
\ee
\noindent
which is independent of the $\ww$ normalization. One finds
\bel{W2beta}
\overline{\ws} = {{\bbar - 1}\over{1 + \beta}} \comma
\ee
\bel{Ebeta}
\ebar = \half {\bbar}^2 - \half {{{(\bbar - 1)}^2}
\over{1 + \beta}} \comma
\ee
\noindent and
\bel{hbeta}
h = 1 + (\bbar - 1){{\beta}\over{1 + \beta}} \period
\ee
Using Eqs.\ (\ref{eq:U}), (\ref{eq:scale}), and (\ref{eq:adef})
we can write
\bel{Vave}
{\overline{V(\vec{\rho}~) \ws}}
= {{{\overline{V(\vec{\rho}~, m_Z ) \ws}}
 - {\overline{V(\vec{\rho}~, m_H ) \ws}}}
\over{\bbar~{\sin}^2 \theta}} \comma
\ee
where
\bel{VrhoM}
V(\vec{\rho}~, m )=
{{1}\over{2 \pi}} \int d^2 \rho'~{{2 \pi k}\over{\ti}}~
K_0 (m \rho' \sqrt{{{2 \pi k}\over{\ti}}})~
{|\ww (\vec{\rho} + {\vec{\rho}} \, ')|}^2
\ee
and
\bel{mx}
m_X = {{1}\over{\sqrt{\bbar}}}{{M_X}\over{M_W}} \hspace{1cm}
(X = Z, H) \period
\ee

For $M_H \geq M_Z$ we have $\beta \geq 0$ and
therefore
$1 \leq h < \bbar$. The uniform field $h$ will remain frozen at
${\bb}_{c2} = 1$ if and only if $M_H/ M_Z = 1$, which is in agreement with
the result of Ref.\,\cite{AO2} for that mass ratio.

This concludes the general
description of electroweak Ginzburg-Landau theory. We have identified
the physical quantities of interest.
The purpose of this paper is to express them, analytically, as a function of
the lattice geometry or, more precisely, of the lattice parameter $\tau$.
In order to do so, we shall have to find a representation of the lattice
solutions that will make possible an evaluation of the integral
 and averages in Eqs.\ (\ref{eq:VrhoM}) and (\ref{eq:Vave}).

\newpage
\section{Expansions of Abrikosov Solutions}
\label{sec-ExpAS}

The Abrikosov flux lattice solutions of \eref{Wlin} were first
derived \cite{Abrikosov} in the gauge
$\vec{\bar{A}} = \bar{B} x \hat{y}$. We prefer the cylindrically
symmetric gauge
$\vec{\bar{A}} =
(\bar{B} \hat{z} \times \vec{r}~)/2$, as it leads to a quicker and more
elegant derivation.

The lowest order perturbative solutions of \eref{Wlin} are
obtained by replacing ${\cal O} (\wsq)$ with an effective mass term
$M_C^2$ (see Ref.\,\cite{MT1992a}).
After transforming to the rescaled units defined by \eref{scale},
and after expressing the vector
$\vec{\rho}~$ in cylindrical coordinates ($\rho$,~$\varphi$),
the equation becomes
\bel{PDE}
\left[
\mbox{} - {{1}\over{\rho}} {{\partial}\over{\partial \rho}} \rho
{{\partial}\over{\partial \rho}}
- {{1}\over{{\rho}^2}} {{{\partial}^2}\over{{\partial \varphi}^2}}
+ 2 i \kappa {{\partial}\over{\partial \varphi}} + {\kappa}^2 {\rho}^2
+ {{2 \kappa}\over{\bbar}}
{\left( 1 + {{M_C^2}\over{M_W^2}} \right)} - 4 \kappa
\right] {\cal W} = 0 \period
\ee
\noindent
It is easily checked that
the angular momentum eigenstates with
eigenvalues $m: m \geq 0$
\bel{fm}
{\ww}_m(\rho ,\varphi) = {{1}\over{\sqrt{\pi m !}}}~
{\kappa}^{{{m+1}\over{2}}} \exp (- {{1}\over{2}} \kappa {\rho}^2 )
{}~{\rho}^m e^{i m \varphi}
\ee
\noindent
and orthonormality condition
\bel{ON}
\int d^2 \rho~{{\ww}_{m'}(\rho ,\varphi)}^\ast~{\ww}_{m}(\rho ,\varphi)
= \delta_{m'm}
\ee
\noindent
are infinitely degenerate solutions of \eref{PDE},
satisfying the eigenvalue relation
\bel{eigen}
\bbar = 1 + {{M_C^2}\over{M_W^2}} \period
\ee
\noindent
If we introduce the complex variable $z = \rho e^{i \varphi}$, the most
general solution is
\bel{Gen}
\ww (\rho , \varphi) = \sum_{m = 0}^{\infty} c_m {\ww}_m(\rho ,\varphi) =
\exp (-{{1}\over{2}} \kappa z z^{\ast}) ~f(z) \comma
\ee
\noindent
where $f(z)$ is an arbitrary analytic function.
\newpage
\noindent
We are interested in solutions with $\ws$ invariant under the lattice
translations $z \rightarrow z + 1$ and $z \rightarrow z + \tau$.
Consider therefore the transformation properties of the
theta function.
\beal{Thetaprop}
\th (\pi (z+1) | \tau) & = & - \th (\pi z | \tau) \comma \nonumber \\*
\th (\pi (z+\tau) | \tau) & = & - \exp ( - i \pi \tau - 2 i \pi z)~
\th (\pi z| \tau) \period
\end{eqnarray}
\noindent
The change of modulus under the second translation can be compensated for
by attaching a prefactor with suitable transformation properties.
If we make the choice
\bel{fanal}
f(z) = {(2 \ti)}^{{{1}\over{4}}}
\exp({{1}\over{2}} \kappa z^2 )
{}~\th (\pi z| \tau)
\ee
\noindent
for $f$ in \eref{Gen}, it is easily seen
that $\ww$ will transform
by at most a phase under the two distinct translations,
provided $\kappa = \pi /\ti$.
According to \eref{scale}, this is
the solution\fotnot{A proof of uniqueness of solutions was provided in
Ref.\,\cite{MT1992a}.}
 corresponding to a single
quantum ($k=1$) of magnetic flux per lattice cell:
\bel{W1}
\ww (\rho , \varphi)
= {(2 \ti)}^{{{1}\over{4}}}
 \exp {\left[{{\pi}\over{2 \ti }} ( z^2 -  z z^{\ast})\right]}
{}~\th (\pi z| \tau) \period
\ee
\noindent
The normalization is chosen so that, as we shall see,
$\ws$ is coordinate covariant under the modular group and the
spatial average $\overline{\ws}$ is equal to one.
The solutions with higher $k$ are given simply by
${[\ww (\rho, \varphi)]}^k$.

Since $\ws$ is doubly periodic, it can be expanded in a Fourier series in
the coordinates $(u, v)$ defined by $z=u + v \tau$.
{}From \eref{W1} and the
series representation
\bel{thseries}
\th (\pi z | \tau) = {{1}\over{i}} \sum_{n = - \infty}^\infty
{(-1)}^n q^{{(n+{{1}\over{2}})}^2} e^{i (2n+1) \pi z} \comma
\ee
\noindent
where $q=e^{i \pi \tau}$, $\ws$ can be written
\bel{W2sum}
{|\ww (u,v)|}^2 = {(2 \ti )}^{{1}\over{2}} \sum_{n=-\infty}^\infty
 ~\sum_{n'=-\infty}^\infty~{(-1)}^{n + n'}
e^{i \pi \tau {\left[v + \left( n + {{1}\over{2}} \right) \right]}^2}
e^{- i \pi \tau^\ast {\left[v - \left( n' + {{1}\over{2}} \right) \right]}^2}
e^{i 2 \pi (n - n') u} \period
\ee
\newpage
\noindent
By trading the dummy index $n$ for $k = n-n'$, the Fourier components in
the $u$ coordinate are already explicit.
Since the expression is not termwise periodic in $v$, integration in $v$
over merely a period will not help. Consider instead the continuous
Fourier transform with respect to the $v$ coordinate,
\beal{Wup}
{|\ww (u;p)|}^2 & = & \int dv~e^{-i 2 \pi p v} {|\ww (u,v)|}^2
\nonumber \\*
& = & \sum_{k= -\infty}^\infty {(-1)}^k e^{i 2 \pi k u}
\exp {\left( -
{{\pi {|k \tau -p|}^2}\over{2 \ti }} \right)}
e^{i \pi p (k+1)} \sum_{n' = -\infty}^\infty e^{i 2 \pi p n'} \period
\end{eqnarray}
\noindent
{}From the inverse Fourier
transform
and
the Poisson formula
\bel{Poisson}
\sum_{n'=- \infty}^\infty~e^{i 2 \pi p n'} =
\sum_{l=- \infty}^\infty \delta (p - l) \comma
\ee
\noindent
one then obtains the result
\bel{Fourier}
{|\ww (u,v)|}^2 = \sum_{k=-\infty}^\infty~\sum_{l=-\infty}^\infty
{(-1)}^{kl + k + l} \exp {\left( -
{{\pi {|k \tau -l|}^2}\over{2 \ti }} \right)}
{}~~e^{i 2 \pi (k u + l v)} \period
\ee

This representation converges extremely fast and uniformly on
the plane.
We remark that the expression is invariant under the
modular group
generated by the transformations
$\tau \rightarrow \tau' = \tau + 1$ and
$\tau \rightarrow \tau' = - 1/ \tau$,
and under reflexion in the imaginary axis
\mbox{$\tau \rightarrow \tau' = - \tau^\ast$}, provided $u$ and
$v$ transform covariantly, \ie $u \rightarrow u'$, $v \rightarrow v'$
where $u + v \tau = u' + v' \tau'$ \cite{MT1992a}.
With the chosen normalization,
 the constant term $1$ can be identified with the average value
$\overline{\ws}$.

The Fourier expansion of $\ws$, \eref{Fourier}, facilitates considerably
 the evaluation
of integrals which occur in expressions for physical quantities
in theories which allow Abrikosov flux
lattice solutions, such as Type-II superconductivity or the electroweak
theory.
In particular it will enable us to write down an analytical expression
for the parameter $\beta$ [see \eref{beta}] in terms of the
lattice parameter $\tau$.

\newpage
\section{Analytical Expressions for Physical Quantities}
\label{sec-AnalExp}

In the lowest order of perturbation theory, physical quantities depend on
the lattice geometry through dimensionless parameters
which are specific to the Abrikosov solutions but independent of
their overall
normalization. In the theory of superconductivity, the
naturally arising geometrical quantity
is the Abrikosov parameter
\bel{betaA}
\beta_A = {{\overline{{|\ww|}^4}}\over{{{( \overline{\ws} )}^2}}}
= \klsum \exp {\left( -
{{\pi {|k \tau -l|}^2}\over{\ti }} \right)} \period
\ee
\noindent
The last equality was obtained by extracting the constant term in the
Fourier expansion of ${|\ww |}^4$ that resulted from squaring
\eref{Fourier}. We remark that the right-hand side of
\eref{betaA}, and in fact any function of ${|k \tau - l|}^2 /\ti$,
summed over all integers $k$ and $l$, is modular invariant.

According to section \ref{sec-EWGL},
the corresponding quantity in the electroweak theory is
\[
\beta = {\textstyle {\overline{V(\vec{\rho}~) \ws}}\over
{\textstyle {( \overline{\ws} )}^2}} \period
\]
The new feature here is the nonlocal quartic interaction
in the numerator.
It is described, as shown in Eqs.\
(\ref{eq:Vave}) and (\ref{eq:VrhoM}), by an integral kernel.

In order to evaluate $\beta$, we first find the Fourier expansion of
the function $V(\vec{\rho}~, m )$ defined by \eref{VrhoM}
and restrict ourselves
to the case $k = 1$. With the representation
\bel{K0}
K_0(x)= \half \int_0^\infty {{dt}\over{t}} \exp
{\left(-t - {{x^2}\over{4 t}} \right)} \comma
\ee
integrations are straightforward, and one finds
\bel{VuvM}
V(u,v~;m) =
\klsum
{(-1)}^{kl + k + l}{\left[ m^2 +
 2 {{\pi {|k \tau -l|}^2}\over{\ti }} \right]}^{-1}
\exp {\left( -
{{\pi {|k \tau -l|}^2}\over{2 \ti }} \right)}
{}~~e^{i 2 \pi (k u + l v)} \period
\ee
\newpage
\noindent
The average $\overline{V(\vec{\rho}~, m ) \ws}$ is
then obtained
by multiplying Eqs.\ (\ref{eq:VuvM}) and (\ref{eq:Fourier})
together and extracting
the constant term.
With $m_Z$ and $m_H$ defined by \eref{mx},
the resulting expression for $\beta$ is
\bel{betatau}
\beta = {{b(\tau, m_Z ) - b(\tau, m_H )}\over{\bbar~{\sin}^2 \theta}}
\comma
\ee
where
\beal{btauM}
b(\tau,m) & = &{\textstyle {\overline{V(\vec{\rho}~,m) \ws}}\over
{\textstyle {( \overline{\ws} )}^2}} \nonumber \\*
& = &
\klsum
{\left[
m^2 + 2 {{\pi {|k \tau -l|}^2}\over{\ti }} \right]}^{-1}
\exp {\left( -
{{\pi {|k \tau -l|}^2}\over{\ti }} \right)} \period
\end{eqnarray}

The quantity $\beta$ depends on, besides
$\tau$, the masses of the two bosons that mediate the interaction and
the redefined magnetic field $\bbar$. It is therefore not a
scale independent geometric parameter in the same sense as the Abrikosov
number $\beta_A$.
The dependence on $\bbar$ enters through the size of the flux lattice,
which becomes significant with the introduction of
interaction scales $M_Z^{-1}$ and $M_H^{-1}$ in the nonlocal kernel.
We can,
however, consistently set $\bbar=1$ in
\eref{betatau}
within the order of perturbation theory we are
considering.

The above calculation can be done also for $k > 1$. Although the results
are not as elegant and multiply convoluted sums abound, considerable
computation time can be gained versus numerical integration. It has
been shown numerically \cite{MT1992a,StJames} that
the solutions with $k>1$ have higher energy, and for this reason
they are not the focus of this paper.

The behavior of $\beta$ in terms of the lattice
parameter $\tau$
has been investigated numerically
in Ref.\,\cite{MT1992a}, where it was found
that, for $M_H > M_Z$,
the global minimum of $\beta$ occurs at
$\tau = e^{i \pi /3}$ corresponding to a
triangular lattice.
With the above results, it is now possible to
show analytically that $\tau = e^{i \pi /3}$ is a local minimum.
\newpage
\noindent
{}From Eqs.\ (\ref{eq:betatau}) and (\ref{eq:btauM}) we have
\bel{betadm}
\beta = {{1}\over{\bbar~{\sin}^2 \theta}}
\int_{m_Z}^{m_H}
{dm~\beta_m} \comma
\ee
where
\bel{betam}
\beta_m \equiv -{{d}\over{dm}} b(\tau ,m) =
\klsum F_m (\osq ) \comma
\ee
\bel{fmx}
F_m (x) = {{2 m}\over{{(m^2 + 2x)}^2}}~e^{-x} \comma
\ee
and the rescaled lattice vectors $\okl{~}$ are given by
$\okl{~} = \sqrt{{{\pi}\over{\ti}}} (k \tau - l)$.
The properties of $\beta$ can be demonstrated by analysis which holds
 true for each $\beta_m$ separately. Introducing
\bel{ddtau}
\dd{}{{\tau}^*} = \half {\left( \dd{}{\tr} + i \dd{}{\ti} \right)}
\hspace{.3cm} , \hspace{.3cm}
\dd{}{\tau} = \half {\left( \dd{}{\tr} - i \dd{}{\ti} \right)}
\ee
one finds
\bel{flat}
\dd{\beta_m}{{\tau}^*} = -{{i}\over{2 \ti}}
\klsum \okl{2}~F_m{}'(\osq ) \period
\ee
For the square $(n = 4)$ and triangular $(n=6)$ lattices with $n$--fold
rotational symmetry and $\tau = e^{i 2 \pi / n}$, all non-zero
lattice vectors $\omega$ appear in $n$-tuplets
$\{\omega {\tau}^r,~r = 1 \ldots n \}$. Then, from the
cancellation of phases within each $n$-tuplet,
the right-hand side of \eref{flat} sums
to zero, and therefore the square and the triangular lattices are stationary
points with respect to the variables $\tr$ and $\ti$.
This could have been shown directly from
modular invariance \cite{M1992b}.

In order to determine the type of local extremum
that the triangular lattice constitutes, we must examine the
eigenvalues of the Hessian
matrix
\bel{hess}
H =
\left(
\begin{array}{l@{\hspace{.2cm}}r}
\dddd{\beta_m}{\tau}{{\tau}^*} & \ddd{\beta_m}{\tau} \\*[.3cm]
\ddd{\beta_m}{{\tau}^*} & \dddd{\beta_m}{\tau}{{\tau}^*}
\end{array}
\right)
= \half U^{\dag}
\left(
\begin{array}{l@{\hspace{.2cm}}r}
\ddd{\beta_m}{\tr} & \dddd{\beta_m}{\tr}{\ti} \\*[.3cm]
\dddd{\beta_m}{\tr}{\ti} & \ddd{\beta_m}{\ti}
\end{array}
\right)U \comma
\ee
where $U$ is a unitary matrix.
\newpage
\noindent
The elements of $H$ are given by
\begin{eqnarray}
\label{eq:h1}
\dddd{\beta_m}{\tau}{{\tau}^*} & = &
{{1}\over{4{\ti}^2}}
\klsum {\left[
2~\osq F_m{}'(\osq ) + {|\okl{ } |}^4 F_m{}''(\osq )
\right]} \comma \\*
\label{eq:h2}
\ddd{\beta_m}{{\tau}^*} & = &
- {{1}\over{4{\ti}^2}}
\klsum {\left[
2~\okl{2}~F_m{}'(\osq ) + \okl{4}~F_m{}''(\osq )
\right]} \comma
\\*
\label{eq:h3}
\ddd{\beta_m}{\tau} & = &
{\left(
\ddd{\beta_m}{{\tau}^*}
\right)}^* \period
\end{eqnarray}
If the summation is performed separately over each $n$-tuplet
$\{\omega e^{i \pi r/3},~r = 1 \dots 6 \}$
 with common modulus, we
see that the right-hand sides of Eqs.\
(\ref{eq:h2}) and (\ref{eq:h3}) are zero by
the cancellation of phases. The Hessian matrix is therefore diagonal with
a double eigenvalue.

To show that the eigenvalue is positive,
examine the expression
$2 x F_m{}'(x) + x^2 F_m{}''(x)$
that occurs
in each term of \eref{h1}. It is easily
shown to be
positive for all $m > 0$, provided $x > 2$. Now for the triangular lattice
we have
\bel{obound}
\osq \geq {{\pi}\over{\ti}} = {{2 \pi}\over{\sqrt{3}}} > 2
\ee
for each non-zero lattice vector $\okl{}$. Therefore, the right-hand side
of
\eref{h1} is positive and, for all $m > 0$, it follows
that $\beta_m$ has a local minimum at $\tau = e^{i \pi /3}$.
{}From \eref{betadm} one then concludes
that $\beta$ has a local minimum for this
value of $\tau$, provided $M_H > M_Z$.

The analysis of the Abrikosov parameter $\beta_A$ can be carried out
similarly.
An investigation of the perturbative properties of theories with
more general quartic interactions
is underway \cite{M1992b}.

According to \eref{Ebeta},
the energy is a monotonically increasing function of $\beta$.
If we assume that $M_H > M_Z$,
the global minimum of $\beta$ occurs at $\tau = e^{i \pi /3}$ \cite{MT1992a},
and it follows that the triangular
lattice solution represents the ground state of
the electroweak vacuum at
magnetic fields above and close to the critical field $B_{c2}$.
\newpage
\vspace{1cm} \centerline{\bf Acknowledgments}
It is a pleasure to thank Samuel MacDowell for many valuable
suggestions.

\end{document}